\documentclass[%
preprint,
amsmath,amssymb,
aps,
prl,
]{revtex4-1}

\usepackage{graphicx}
\usepackage{dcolumn}
\usepackage{bm}

\begin{document}

\preprint{APS/123-QED}

\title{Phonon Spectroscopy with Sub-meV Resolution by Femtosecond X-ray Diffuse Scattering}

\author{Diling Zhu$^1$}
\author{Aymeric Robert$^1$}
\author{Tom Henighan$^{2,3}$}
\author{Henrik T. Lemke$^1$}
\author{Matthieu Chollet$^1$}
\author{J.~Mike Glownia$^1$}
\author{David A. Reis$^{2,4,5}$}
\author{Mariano Trigo$^{2,4}$}\thanks{mtrigo@slac.stanford.edu}

\affiliation{$^1$Linac Coherent Light Source, SLAC National Accelerator Laboratory, Menlo Park, California, USA}

\affiliation{$^2$Pulse Institute, SLAC National Accelerator Laboratory, Menlo Park, California, USA}

\affiliation{$^3$Physics Department, Stanford University, Stanford, California, USA}

\affiliation{$^4$SIMES Institute, SLAC National Accelerator Laboratory, Menlo Park, California, USA} 

\affiliation{$^5$Department of Photon Science and Applied Physics, Stanford University, Stanford, California, USA}

\date{\today}

\begin{abstract}
We present a reconstruction of the transverse acoustic phonon dispersion of germanium from femtosecond time-resolved x-ray diffuse scattering measurements at the Linac Coherent Light Source. We demonstrate an energy resolution of $0.3$~meV with momentum resolution of $0.01~$nm$^{-1}$ using $10$~keV x-rays with a bandwidth of $\sim 1$~eV. This high resolution was achieved simultaneously for a large section of reciprocal space including regions closely following three of the principle symmetry directions. The phonon dispersion was reconstructed with less than three hours of measurement time, during which neither the x-ray energy, the sample orientation, nor the detector position were scanned. These results demonstrate how time-domain measurements can complement conventional frequency domain inelastic scattering techniques.
\end{abstract}

\pacs{xxx.xxx}

\maketitle

Inelastic neutron and x-ray scattering have contributed greatly to our knowledge of collective excitations and dynamics in condensed matter~\cite{Squires1978,Burkel1991}. Such energy-loss measurements are normally performed in the frequency domain where the energy lost by the scattered pseudo-particle is resolved by a crystal analyzer. In the case of inelastic x-ray scattering (IXS), heroic efforts have been made in the development of high resolution monochromators and analyzers to achieve the resolution needed to reveal the nature of low energy excited states~\cite{Sette1995,Masciovecchio1996,Sette1998,Baron2001,Sinn2001,Shvydko2014,Ishikawa2015}. Inevitably, higher energy and momentum resolution are achieved at a price of reduction in count rate as well as increase in the size and complexity of the instrumentation. Recent advances in hard x-ray free electron laser (FEL) sources~\cite{Emma2010,Ishikawa2012} provide femtosecond x-ray pulses with sufficient intensity for capturing density fluctuations directly in the time domain. Thus, provided the collective modes can be excited coherently, one can directly measure their frequency as a function of momentum via the Fourier transform (FT) of the observed time-dependent scattered intensity, as first demonstrated by Trigo et al.~\cite{Trigo2013}. An advantage of this approach is that it does not require high-resolution monochromators or spectrometers. Therefore it benefits from much higher incident flux on the sample by using radiation of a bandwidth much wider than the targeted excitation energy. In addition, the use of a 2D detector enables the simultaneous coverage of very large regions of reciprocal space. 

\begin{figure}
\centering
\includegraphics[width=3.2in]{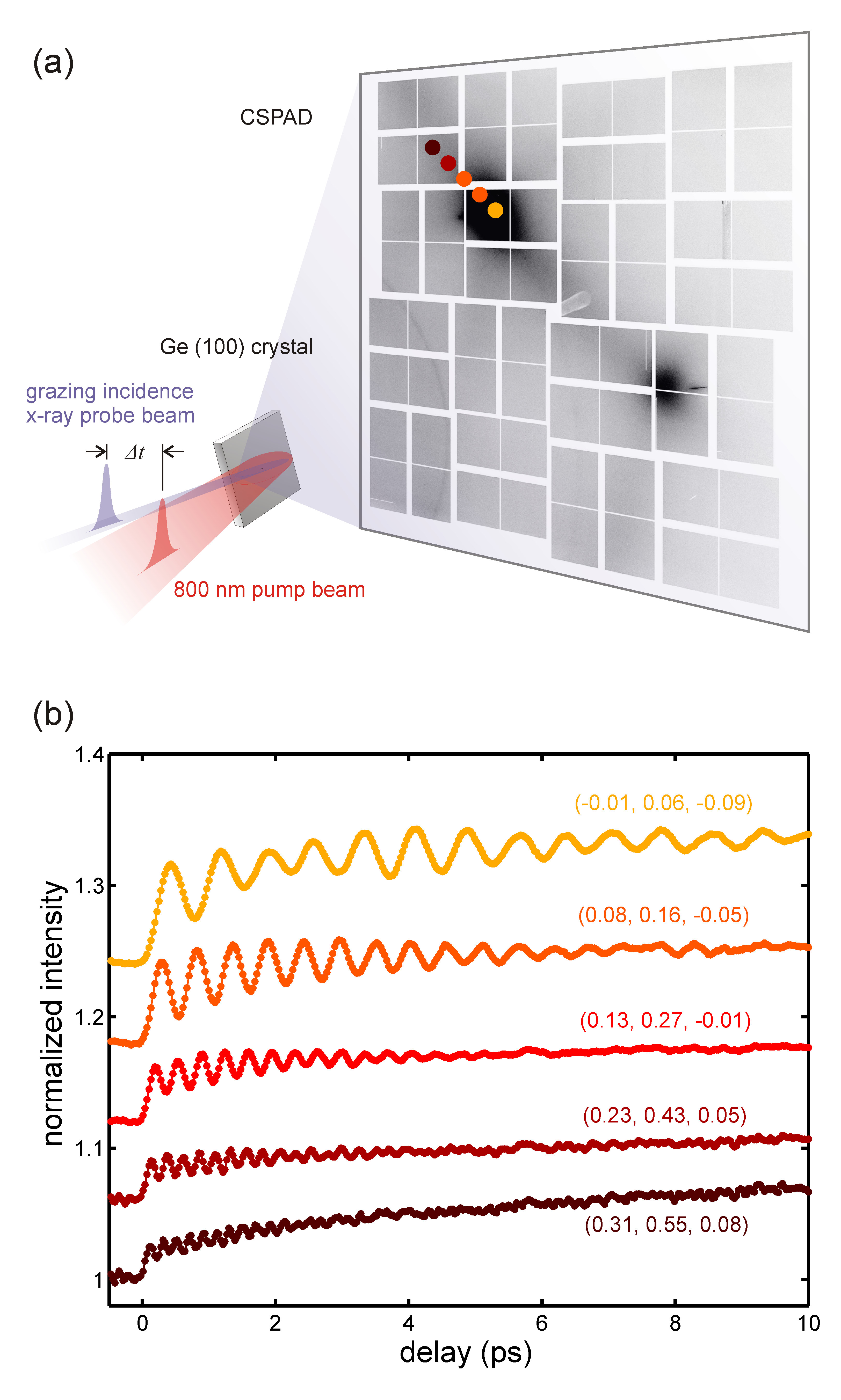}
\caption{(a) Schematics of the grazing incidence pump-probe setup. A multi-shot averaged thermal diffuse scattering pattern is shown in gray scale. Five distinct locations are indicated by the colored dots on the CSPAD. (b) The time evolution of the normalized scattering intensity as a function of pump-probe delay (at the five locations indicated by the corresponding colored dots on the scattering pattern in (a)) is shown using the corresponding color scheme. The reduced $(hkl)$ positions are indicated for each time trace.}
\label{fig:f1}
\end{figure}

In a previous work~\cite{Trigo2013} we reported the observation of coherent phonons with energies between 2 and 5.4 meV using this approach. The limits to the accessible frequencies were suggested to be due to beam fluctuations of the FEL and to the arrival time between the pump and the probe beams. In this letter we present optical pump x-ray probe measurements using the improved capabilities of the X-ray Pump Probe (XPP) instrument~\cite{Chollet2015} at the Linac Coherent Light Source (LCLS). Using germanium as a benchmark, we identify the fundamental aspects that impact the frequency and momentum resolution of this technique. This experiment improves significantly beyond the initial results in~\cite{Trigo2013} with the addition of the shot-by-shot correction of the pump-probe timing jitter and partial monochromatization of the x-ray probe. The higher time resolution allows us to obtain the spectrum of transverse acoustic (TA) phonons on a frequency range over twice as large and much better resolution than previously reported. We were able to map a large portion of the TA phonon dispersion of germanium over an extended two dimensional section of the reciprocal space spanning two Brillouin zones (BZs), reaching some of the high-frequency TA modes at the edge of the zone. Our results also confirm that the limited frequency range observed in~\cite{Trigo2013} was caused by the instrument time resolution rather than any intrinsic physical limitation.



The measurement was performed in the large-offset double crystal monochromator mode of the XPP instrument with the Si(111) configuration~\cite{Zhu2014}, leading to a bandwidth of 1.3 eV at 10.2 keV. A germanium single crystal of the (100) orientation, 10$\times$10~mm$^2$ in size, was used for the measurement. The x-ray beam was focused to produce a spot size of $50\times 50~\mu$m$^2$ full width at half maximum (FWHM) at the sample with a 0.5 degree grazing incidence angle. This incidence angle was chosen to match the optical and x-ray penetration depths. An 800 nm Ti:Sapphire pump laser with 50 fs pulse duration was used as the optical excitation focused to a $400\times400~\mu$m$^2$ FWHM. The optical pulse energy was 400 $\mu$J, and was $p$-polarized. The laser incidence angle was 1.5 degree grazing from the surface, leading to a pump incidence fluence of 6.5 mJ$\cdot$~cm$^{-2}$. The small angle between the pump and probe incident directions means that the differential arrival time between pump and probe across the beam footprint is negligible ($<$10~fs). The spectro-encoding technique was used for pulse-to-pulse timing correction~\cite{Harmand2012,Lemke2013b,Bionta2014}. A 15~$\mu$m thick Ce:YAG crystal was used as the timing-cross-correlation target. The x-ray pulse duration was $\sim$ 50 fs based on electron bunch length measurement. With an estimate of 25~fs precision of the timing-diagnostic, the overall time resolution of the experiment is better than 80~fs. The setup utilized a vacuum chamber to minimize  background scattering. The scattered x-rays were recorded using the 2.4 megapixel CSPAD detector~\cite{Hart2012,Blaj2015} placed 195 mm downstream the sample. The sample azimuth angle was adjusted such that the Ewald's sphere closely cut through the (022) and (113) Brillouin zones (indices given in the conventional Bravais lattice). A diagram of the sample geometry and the measured static diffuse scattering pattern are shown in Fig.~\ref{fig:f1}(a). The observed static diffuse scattering intensity is mostly from  thermally populated phonons~\cite{Trigo2010,Trigo2013}.

Ultrafast excitation produces correlated pairs of phonons at equal and opposite momenta (${\bf q}$ and ${\bf -q}$), which modulate the scattered intensity $I({\bf G+q}, t)$ at twice the corresponding phonon frequency $2 \omega({\bf q})$~\cite{Trigo2013}, where ${\bf Q} = {\bf G+q}$ is the total momentum transfer, ${\bf G}$ is a reciprocal lattice vector, and ${\bf q}$ is the reduced wavevector.
Figure~\ref{fig:f1}(b) shows the time evolution of the measured scattering intensity $I({\bf Q}, t)$ for the five pixel locations as indicated in the corresponding color in Fig.~\ref{fig:f1}(a). Their corresponding ${\bf q}$ locations are marked next to the time traces. As the corresponding reduced wavevector moves from the zone center (yellow, top) to the zone edge (dark red, bottom), the frequency of the oscillations increases from approximately 1~THz to approximately 5 THz, reflecting the dispersion of twice the transverse-acoustic (TA) phonons. Such long-lived oscillatory dynamics was observed across the entire detector, which spans the entire (022) and (113) BZs.

\begin{figure*}
\centering
\includegraphics[width=6.5in]{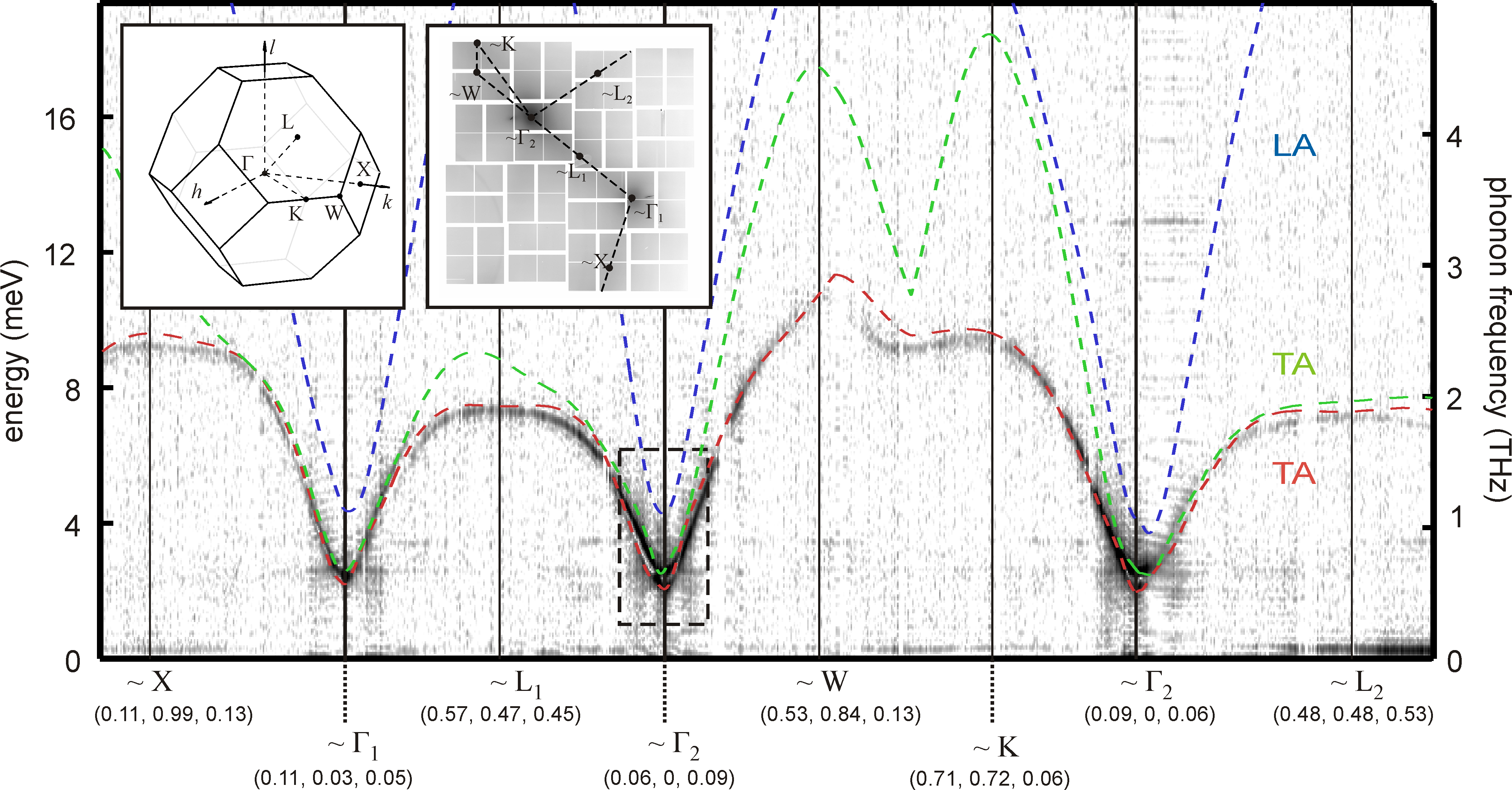}
\caption{Reconstruction of the phonon dispersion relation along the path in reciprocal space indicated in the right inset. Locations of this path close to high-symmetry points in the Brillouin Zone (BZ) are labeled with their respective wavevector in reciprocal lattice units (r.l.u.). The left inset shows the BZ with the high symmetry points. The right inset shows the multi-shot averaged diffuse scattering pattern and superimposed  path ( as indicated by the dashed line) taken to obtain the main plot. The dashed box near $\Gamma_2$ in the main plot indicates the region that is magnified in Fig.~\ref{fig:f3}(a).
}
\label{fig:f2}
\end{figure*}

Figure~\ref{fig:f2} shows the extracted TA dispersion $\omega({\bf q})$ along the reciprocal space path as indicated in the insets. This path cuts the BZs near some of the main symmetry points. Due to limited detector dynamic range, the geometry was intentionally chosen to avoid the Bragg condition, such that the reciprocal space section captured by the detector bypassed the Gamma points. This leads to the lower limit of the frequency modes that are measured. The grey scale represents the FT amplitude (logarithmic scale) of the oscillatory component of $ I({\bf Q},t)$, and the phonon frequency axis is obtained by dividing the FT frequency by two. The dashed lines in Fig.~\ref{fig:f2} show the three acoustic branches, the two transverse and one longitudinal, calculated using a harmonic model of the forces, using six nearest neighbors~\cite{Herman1959}. Note that only modes with a component of the polarization along the scattering vector are expected to contribute to first order diffuse scattering, therefore primarily one branch is visible in this geometry~\cite{Trigo2013}. 

\begin{figure}
\centering
\includegraphics[width=2.5in]{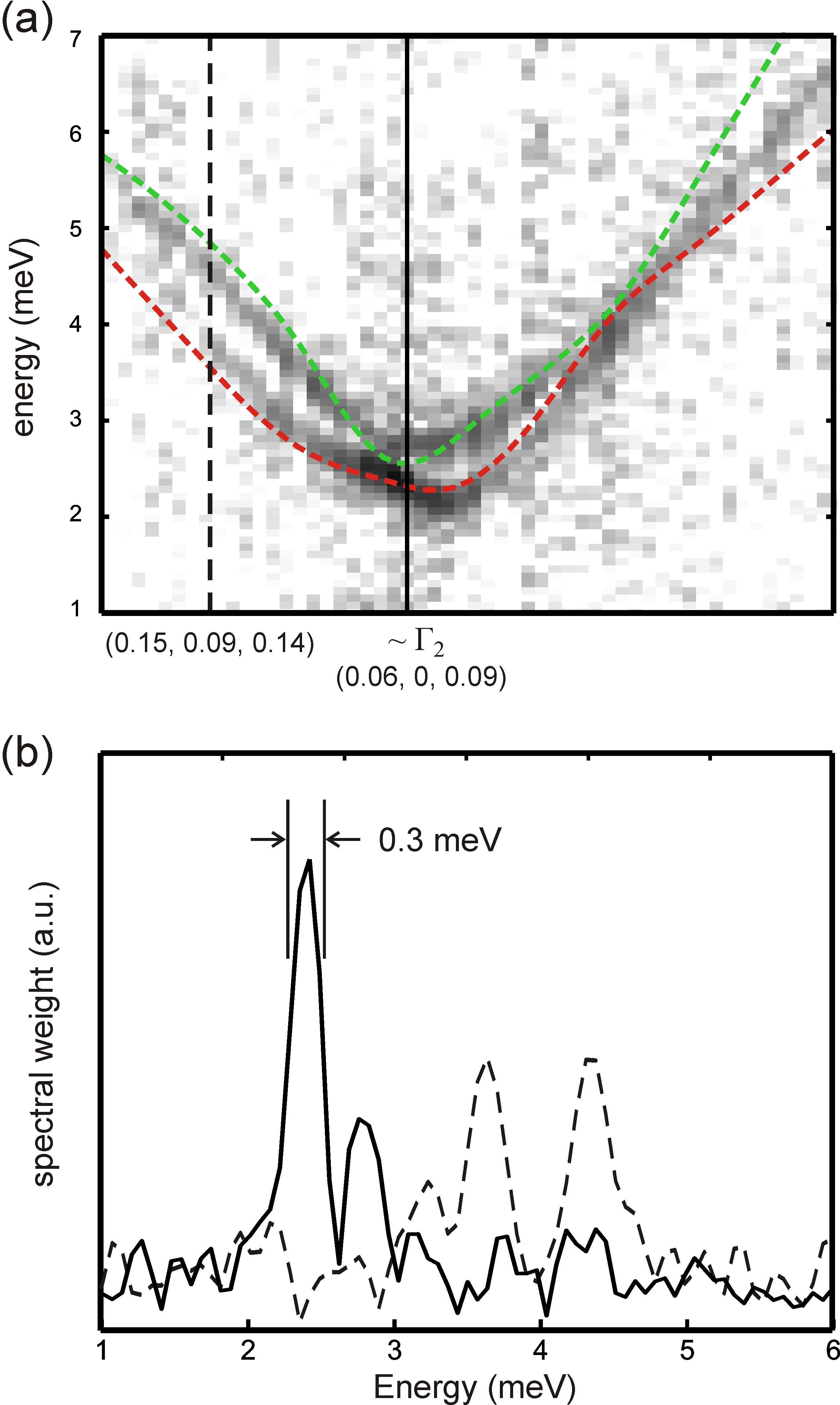}
\caption{(a) Detailed view of the dispersion relation  presented in the dashed box in Fig.~\ref{fig:f2}. This was obtained from a single delay scan over 8 minutes. (b) Energy lineouts for the two wavevectors as indicated in (a) by the dashed and solid lines.}
\label{fig:f3}
\end{figure}

To highlight the momentum and energy resolution, Fig.~\ref{fig:f3}(a) shows an expanded view of the dispersion in the region near the BZ center (region indicated by the dashed rectangle in Fig.~\ref{fig:f2} near $\Gamma_2$). The two separate TA branches are clearly visible. Fig.~\ref{fig:f3} (b) shows the spectrum from (a) at the two wavevector locations indicated by the vertical solid and dashed lines. Near the degenerate point, the two well-resolved TA peaks indicate a resolution better than $\sim 0.4$~meV. The FWHMs of the peaks are as small as 0.3 meV, consistent with being limited by the largest delay time reached by the time scans (i.e. 10ps as shown in Fig.~\ref{fig:f1}(b)).\\

As in any scattering experiment, the momentum resolution $\Delta Q$ is defined in part by the scattering geometry. In our experiment, the main contribution to $\Delta Q$ is from the elongated footprint of the X-ray beam on the sample ($\sim$ 2 mm) as a result of the grazing incidence. This only affects the radial direction on the detector, yielding $\Delta Q = 0.3~{\rm nm}^{-1}$ near the center of the detector. On the other hand, perpendicular to the radial direction, we estimate $\Delta Q=0.01~{\rm nm}^{-1}$, which is a result of the transverse beam size (50~$\mu$m) and the detector pixel size (110~$\mu$m). In contrast to previous results~\cite{Trigo2013}, the use of the Si(111) monochromator eliminated the shot-to-shot x-ray energy fluctuations, thus Ewald sphere radius variations. The incoming x-ray bandwidth of 1.3 eV and the x-ray beam divergence of 100~$\mu rad$ due to focusing both contribute to $Q$ smearing well below $0.01~{\rm nm}^{-1}$.

A finite $\Delta Q$ directly leads to a decrease in frequency resolution $\Delta\omega$ as a result of the phonon dispersion. Perpendicular to the radial direction and near the BZ center, we get $\Delta\omega \sim 0.03$ and $0.05$~THz, or $\sim 0.1$ and $\sim 0.2$~meV for the TA and LA branches respectively, based on the speed of sound of approximately 3000 ms$^{-1}$ and 5000 ms$^{-1}$. On the other hand, looking radially on the detector and using the LA speed of sound we obtain a frequency resolution of 1 THz or 4~meV. Note that because of selection rules, the radial direction is where the scattering is more sensitive to LA modes, which could explain why we only observe the TA branches: the oscillations may be washed out. 

$\Delta Q$ (and the corresponding  $\Delta\omega$) could be reduced by either reducing the beam footprint or increasing the sample-detector distance. One desired feature of our approach is that if no other experimental broadening to the resolution exists, the ultimate limit to the frequency resolution is given by the Nyquist theorem $\Delta \omega_N = 1/(2T)$, where $T = 10$~ps is the total length of the time scan. Thus, one can increase the resolution arbitrarily by simply scanning over longer delays. In the present case we obtain $\Delta \nu_N = 0.05$~THz $ = 0.2$~meV.

The sensitivity and measurement efficiency of this technique can be also seen from Fig.~\ref{fig:f3}, which was obtained with a single scan of 8 minutes including the idle time between scan steps. This is clearly sufficient for resolving the TA branch splitting and resolving details which are otherwise only accessible at state-of-the-art IXS facilities~\cite{Shvydko2014,Ishikawa2015}. We note that since the pump-probe delay is scanned electronically, the dispersion curves in Fig.~\ref{fig:f2} were obtained with no moving motors. Finally, we also note that this method does not require  high-resolution analyzers nor monochromators and thus benefits from orders of magnitude higher signal.


In summary, we have shown that Fourier transform inelastic x-ray scattering is an effective way to measure phonon dispersions in solids, particularly for the low frequency modes. We measured simultaneously a large portion of TA phonon dispersion of germanium across extended areas of reciprocal space. We show that the momentum and frequency resolution are mostly determined by the scattering geometry and the maximum delay. The former can be improved by adjusting the beam footprint on the sample and the detector distance, while the latter simply requires measuring time-resolved signals to longer delays. We note that shorter pulse duration for both the x-ray and the optical pump~\cite{Ding2009,Hartmann2014}, lower noise higher dynamic range x-ray detectors~\cite{Blaj2015}, and higher repetition rate X-ray FELs that will be available in the near future,  could yield orders of magnitude improvements in resolution and sensitivity. Finally, this approach presents a unique combination of spectral resolution and sensitivity to non-equilibrium dynamics, and makes this method ideal for probing the microscopic details of energy transport in nanoscale thin films and heterostructures. This technique may even be extended to the measurement of individual nanostructures~\cite{Clark2013,Raubenheimer2014}, which is currently a big challenge for even the most state-of-the-art IXS instruments.

The experiment was carried out at the LCLS at SLAC National Accelerator Laboratory. LCLS is an Office of Science User Facility operated for the U.S. Department of Energy Office of Science by Stanford University. M.T., T.H., and D.R. were supported by the US Department of Energy (DOE), Office of Basic Energy Sciences (BES) through the Division of Materials Sciences and Engineering under contract DE-AC02-76SF00515.

\end{document}